\documentclass[aps,floatfix,showpacs,preprint]{revtex4}
\usepackage{graphicx,bm}

\begin{document}
\title{Squeezing and entangling nuclear spins in helium 3}

\author{G. Reinaudi, A. Sinatra}
\affiliation{LKB, ENS,
 24 Rue Lhomond, 75231 Paris Cedex 05, France}
\author{A. Dantan, M. Pinard}
\affiliation{LKB, UPMC,
case 74, place Jussieu, 75252 Paris Cedex 05, France}

\begin{abstract}
We present a realistic model for transferring the squeezing or the
entanglement of optical field modes to the collective ground state
nuclear spin of $^3$He using metastability exchange collisions. We
discuss in detail the requirements for obtaining good quantum
state transfer efficiency and study the possibility to readout the
nuclear spin state optically.
\end{abstract}


\maketitle
\section{Introduction}
Helium 3 atoms in their ground state possess a purely nuclear spin
$I=1/2$. Such spins are well-isolated from the environment and
show extremely long coherence times. Longitudinal coherence times
$T_1$ of several days are measured in room temperature samples
\cite{Heil}. The transverse coherence time $T_2$, which would be
as long as $T_1$ in zero magnetic field, is usually limited by
magnetic field inhomogeneity if no special precaution is taken.
Transverse coherence times of several hours are observed in a very
low field \cite{Cohen}. These very long coherence times originate
from the weakness of magnetic coupling on the one hand, and from
the absence of electrical coupling on the other hand, as there is
no electric quadrupole coupling within the ground state for spins
1/2. It is tempting to exploit such long-lived coherence for
quantum information purposes. In a previous letter \cite{prl} we
studied the possibility to transfer the squeezing of a cavity mode
to $^3$He nuclear spins. We showed that the squeezeing could be stored 
and retrieved from the atoms, thus realizing a quantum memory 
\cite{Fleishhauer,molmer,dantanpra04,vanderWal,Schori,Polzik}. For
the sake of simplicity we presented in our letter a simplified
model involving only two sublevels in the metastable state
and gave numerical
results for the more complicated case of $^3$He. In this paper we
concentrate on $^3$He and treat in detail this more realistic
case.

Section \ref{sec:exchange} is devoted to metastability exchange
collisions. We derive linearized Heisenberg-Langevin equations describing
the exchange collisions from the standpoint of quantum fluctuations. 
In section \ref{sec:model} we
describe the model for squeezing transfer from a squeezed vacuum
mode of the electromagnetic field to the atoms. Numerical results
are shown and discussed in section \ref{sec:numerics}. In section
\ref{sec:analy} we obtain analytical results in the adiabatic
elimination limit for the optical coherences and the cavity field.
Section \ref{sec:lecture} is devoted to the readout scheme of the
quantum memory. In section \ref{sec:2cells}, as a straightforward
application of our scheme, we consider the possibility of creating
long-lived quantum correlations between two macroscopic spins, in
the move of the successful experiment in Copenhagen
\cite{Polzik2}, in which two macroscopic spins were entangled for
$0.5$ ms, but on a completely different timescale. Finally, in
section \ref{sec:Pdiff1}, we use a toy model to explore the
consequences of an imperfect polarization of the atoms on our
squeezing transfer scheme.

\section{Metastability exchange collisions in helium 3}
\label{sec:exchange}

Over forty years ago, Colegrove, Schearer and Walters \cite{CSW}
demonstrated a technique to polarize $^3$He relying on
$(i)$ an optical interaction on an infrared transition from the
metastable $2^{3}$S triplet state to the $2^{3}$P triplet state,
and $(ii)$ \textit{metastability exchange collisions} between
atoms in the ground state and in the metastable state. During such
a collision, two atoms exchange their electronic degrees of
freedom so that the metastable atom, oriented by optical pumping
and with a nuclear polarization due to hyperfine coupling in the
metastable state, becomes a polarized ground state atom
\cite{partridge}. This technique called \textit{metastability
exchange optical pumping} is currently used to prepare polarized
samples for nuclear physics experiments as well as in nuclear
magnetic resonance imaging applications \cite{LeducReview}.

In what follows we suggest that metastability exchange collisions
can also be used to transfer quantum correlations to the ground
state nuclear spin of $^3$He.

\subsection{Equations for the one-body density matrix elements}

Partridge and Series \cite{partridge} describe a metastability
exchange (ME) collision in terms of the one-body density matrices
representing the internal states of two colliding atoms that we
name $\rho^{\rm at}_{\rm g}$ and $\rho^{\rm at}_{\rm m}$ for the
ground and metastable state, respectively. The density matrices
after the collision, ${\rho^{\rm at}_{\rm g}}'$ and ${\rho^{\rm
at}_{\rm m}}'$, are given by
\begin{equation}\label{eq:rho_prime}
\left\{
    \begin{array}{lcl}
      {\rho^{\rm at}_{\rm g}}' &=& \textrm{Tr}_e \;\rho^{\rm at}_{\rm m} \\
      {\rho^{\rm at}_{\rm m}}' &=& \rho^{\rm at}_{\rm g} \otimes
\textrm{Tr}_n\; {\rho^{\rm at}_{\rm m}} \\
    \end{array}
\right.
\label{eq:series}
\end{equation}
where $\textrm{Tr}_e$ and $\textrm{Tr}_n$ are trace operators over
the electronic and nuclear variables.  

Let us consider $n$ metastable and $N$ ground state independent atoms.
We introduce $\rho_{\rm g} = N \rho^{\rm at}_{\rm g}$, $\rho_{\rm m} = n
\rho^{\rm at}_{\rm m}$, and the same for $\rho_{\rm g}^\prime$ and 
$\rho_{\rm m}^\prime$.

To represent the state of the system, we will use the 
density matrix $\rho$ defined by:
\begin{equation}
    \rho =
    \left(%
\begin{array}{cc}
  \rho_{\rm m} & 0 \\
  0 & \rho_{\rm g} \\
\end{array}%
\right)
\end{equation}
Note that $Tr{\rho}=n+N$ and that we neglect all
coherences between the ground and the metastable states.
The matrices $\rho_{\rm g}$ and $\rho_{\rm m}$ evolve according to
\begin{equation}\label{bilan}
\left\{
    \begin{array}{lcl}
        \frac{d }{dt}\rho_{\rm g} &=& - \gamma_f \rho_{\rm g} +
        \gamma_f \rho_{\rm g}^\prime \\
        \frac{d }{dt}\rho_{\rm m} &=& - \gamma_m \rho_{\rm m} +
        \gamma_m \;\rho_{\rm m}^\prime 
    \end{array}
\right.
\label{eq:bilan}
\end{equation}
where $\gamma_f$ and $\gamma_m$ are the metastability exchange
collision rates in the ground and metastable states respectively 
\begin{equation}
\gamma_m= N \gamma_{exc} \hspace{1cm} \gamma_f= n \gamma_{exc}
\end{equation}
with $\gamma_{exc}$ a rate depending on the
metastability exchange cross section, the relative velocity of the
atoms and the volume explored by the atoms.

The calculation of $d\rho/dt$ is performed by expressing $\rho$ in the
decoupled spin basis of the nuclear spin $I=\frac{1}{2}$ and
the total electronic spin $J=S=1$ in the metastable state,
followed by a projection onto the hyperfine states
(eigenstates of the total momentum operator $F$
and $F_z$) which we name from $1$ to $6$ as in figure \ref{fig:schemaMEC}.
The explicit evolution equations for the density matrix elements are
given in the Appendix. The fully polarized state in which all the
atoms are in the sublevel with highest angular momentum projection
along $z$ is stationary for equations (\ref{eq:bilan}).

\begin{figure}[htb]
\centerline{\includegraphics[width=8cm,clip=]{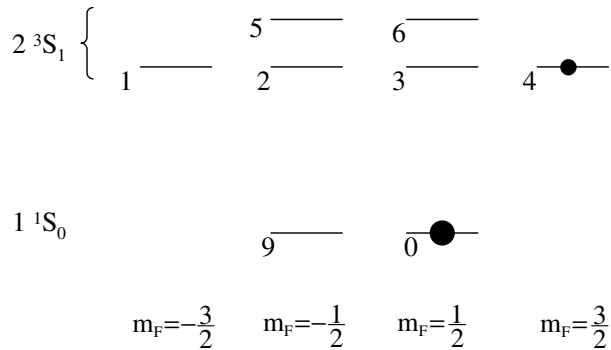}}
\caption{Sublevels $1$ to $6$ are metastable; $9$ and $0$ are
the ground state sublevels. The fully polarized stationary state is shown.}
\label{fig:schemaMEC}
\end{figure}

Starting from equations (\ref{eq:bilan}) we
proceed in two steps which will be detailed in the following:
\begin{enumerate}
\item We linearize these
equations around the fully polarized steady state in which the
only non-zero elements of $\rho$ are $\rho_{44}=n$ and
$\rho_{00}=N$.
\item From the linearized classical equations, interpreted as
semiclassical equations for the mean values of the collective
operators, we derive the corresponding Heisenberg-Langevin
equations.
\end{enumerate}

\subsection{Linearized Heisenberg-Langevin equations}
By linearization around the fully polarized solution we obtain
equations for the ``fluctuations" or deviations of the $\rho_{ij}$
from their steady-state values. Such linear equations coincide
with the linearized semiclassical equations for the collective
atomic operators operators mean values:
\begin{eqnarray}
\rho_{kl}&=& \langle S_{lk} \rangle \hspace{2cm} k,l=1,\ldots,6 \\
\rho_{kl}&=& \langle I_{lk} \rangle \hspace{2cm} k,l=9,0
\end{eqnarray}
where $S_{kl}= \sum_{i=1}^n |k\rangle_i \langle l|_i$ for
$k,l=1,\ldots,6$ and $I_{kl}= \sum_{i=1}^N |k\rangle_i \langle l|_i$
for $k,l=9,0$ are the collective atomic operators in the
metastable and ground state, respectively. The corresponding
linearized Heisenberg-Langevin equation for the operators are
obtained by adding zero-mean valued fluctuating terms which are
the Langevin forces. Denoting by $f_\alpha$ the Langevin force for
the operator $\alpha$ we get a closed set of equations:
\begin{eqnarray}
\dot{S}_{21}&=&-\gamma_m S_{21} + f_{21} \label{eq:uno_exc} \\
\dot{S}_{32}&=& \frac{2}{9} \gamma_m \left(-\frac{7}{2} S_{32} +
    \sqrt{3} S_{21} + S_{65} \right) + f_{32} \\
\dot{S}_{65}&=&-\frac{7}{9}\gamma_m \left(S_{65}-\frac{2}{7}\sqrt{3} S_{21}
-\frac{2}{7} S_{32}\right)  + f_{65} \label{eq:tre_exc} \\
\dot{S}_{43}&=& \gamma_m\left(-\frac{1}{3} S_{43}
    + \frac{2 \sqrt{3}}{9} (S_{32} + S_{65}) \right) +
    \frac{\sqrt{3}}{3} \gamma_f I_{09} +f_{43} \\
\dot{I}_{09}&=&\frac{1}{3}\left[-3 \gamma_f I_{09} +2\gamma_m
\left( S_{32}-\frac{1}{2} S_{65}+\frac{\sqrt{3}}{2}S_{43}+
\frac{\sqrt{3}}{2} S_{21} \right)\right]  +f_{09}
    \label{eq:cinque_exc}
\end{eqnarray}
If $\alpha$ and $\beta$ denote
two system operators, $\langle f_\alpha(t) f_\beta(t^\prime)
\rangle = D_{\alpha \beta } \delta(t-t^\prime)$ where
$D_{\alpha,\beta}$ is the corresponding coefficient of the
diffusion matrix which can be calculated using the generalized
Einstein relations \cite{CohenBook} for an ensemble of uncorrelated
atoms. The non-zero coefficients are
\begin{equation}
\begin{tabular}{lll}
$D_{43,34}=\frac{2}{3}\gamma_m n$, \hspace{2cm}&
$D_{09,34}=D_{43,90}=-\frac{2\sqrt{3}}{3}\gamma_m n$,
\hspace{2cm}& $D_{09,90}= 2 \gamma_m n$.
\end{tabular}
\label{eq:diff_exc}
\end{equation}
Langevin forces are necessary to the consistency of the model.
Otherwise, the non-Hamiltonian character of the exchange terms
leads to a violation of the Heisenberg uncertainty relations.
Physically, these forces originate from the fluctuating character
of the ME collisions and their correlation time is the collision
time, much shorter ($\sim 10^{-13}$ s) than all the times scales
we are interested in.

\subsection{Consequences of the Heisenberg-Langevin equations
for ME collisions}

We notice that Eqs.~(\ref{eq:uno_exc})-(\ref{eq:tre_exc}) for
$S_{21}$, $S_{32}$, $S_{65}$ form a closed subset of equations.
This means that in the frequency domain each of these variables
can be expressed as a linear combination of the Langevin forces
$f_{21}$, $f_{32}$, $f_{65}$. However, in the fully polarized
limit we consider here, these Langevin forces do not contribute to
the diffusion matrix. It follows that these variables do not
contribute to the spin noise and can be neglected. One is then
left with only two equations
\begin{eqnarray}
\dot{S}_{43}&=& -  \frac{\gamma_m}{3} S_{43}
    + \frac{\sqrt{3}}{3} \gamma_f I_{09} +f_{43}
\label{eq:simple_S} \\
\dot{I}_{09}&=& - \gamma_f I_{09} +
    \gamma_m \frac{\sqrt{3}}{3} S_{43}+ f_{09} \,.
\label{eq:simple_I}
\end{eqnarray}
Let us introduce the transverse spin quadratures $S_x$, $S_y$
\begin{equation}
S_x=(S_{34} + S_{43})/2, \hspace{1cm} S_y=i(S_{34} - S_{43})/2
\end{equation}
(and similarly for the ground state spin transverse components 
$I_x$, $I_y$) and let us assume that the
ground state is initially squeezed, while the metastable atoms are
in a coherent spin state. Integrating
(\ref{eq:simple_S})-(\ref{eq:simple_I}) with the initial
conditions $\overline{\Delta I_y^2}(0)=\Delta I_y^2(0)/(N/4)=e^{-2r}$
and $\overline{\Delta S_y^2}(0)=\Delta S_y^2(0)/(n/4)=1$ one finds the
normalized steady state variances to be
\begin{eqnarray}
\overline{ \Delta {S}_y^2}
&=&1-[1-e^{-2r}]\frac{3nN}{(3n+N)^2} \label{eq:corr_S}\\
\overline{ \Delta {I}_y^2}
&=&1-[1-e^{-2r}]\frac{N^2}{(3n+N)^2}
\label{eq:corr_I}
\end{eqnarray}
Since $n\ll N$ (typically $n/N\sim 10^{-6}$), the ground state
spin is still squeezed by approximately the same factor $e^{-2r}$,
whereas the metastable atoms squeezing is negligible (in $n/N$).
By introducing the correlation functions ${\cal C}_S$ and
${\cal C}_I$ of two individual spins in the metastable and ground state 
respectively:
\begin{equation}
{\cal C}_S= \frac{\overline{\Delta {S}_y^2}-1}{4n} \hspace{1cm}
\mbox{and} \hspace{1cm}
{\cal C}_I= \frac{\overline{\Delta {I}_y^2}-1}{4N}
\end{equation}
this simple calculation shows that ME collisions tend to equalize
the correlation function (up to some numerical constant):
$C_S=3C_I$. If the ground state spin is squeezed, $C_I$ has a
negative value of order $1/N$, corresponding to significant collective
correlations for the $N$-particle ensemble. However, as $n \ll N$,
this negative value of the correlation function in the metastable
state is by far too small to induce squeezing into the
$n$-particle metastable state, which would require ${\cal C}_S\sim
-1/n$. For $e^{-2r}=1$ we recover the coherent spin
state with no correlation between the ground state and the
metastable spins.

Noise spectra can also be derived in a similar fashion. By
defining the noise spectrum as
\begin{equation}
{\cal S}_{x_i x_j}(\omega) = \int d\tau \; e^{-i\omega \tau}
\langle x_i(0) x_j(\tau) \rangle
\end{equation}
where $x_i,x_j$ are fluctuations of system operators and
for the same initial conditions $\overline{\Delta {I}_y^2}(0)=\Delta
I_y^2(0)/(N/4)=e^{-2r}$ and $\overline{\Delta {S}_y^2}(0)=\Delta
S_y^2(0)/(n/4)=1$ we get:
\begin{eqnarray}
{\cal S}_{I_y,I_y}(\omega)&=& \frac{\pi (N e^{-2r} + 3n)N^2
\delta(\omega)}{2(N+3n)^2} + \frac{9 \gamma_{\rm exc}nN}{18
\omega^2+2(N+3n)^2 \gamma_{\rm exc}^2}
\label{eq:spectre_I} \\
{\cal S}_{S_y,S_y}(\omega)&=&
\frac{3\pi (N e^{-2r} + 3n)n^2 \delta(\omega)}{2(N+3n)^2}
+ \frac{3 \gamma_{\rm exc}nN}{18 \omega^2+2(N+3n)^2 \gamma_{\rm exc}^2}
\label{eq:spectre_S}
\end{eqnarray}
The equal time correlations (\ref{eq:corr_S}) and
(\ref{eq:corr_I}) can be recovered from these formulas by
integration:
\begin{equation}
\langle x_i x_j \rangle = \frac{1}{2\pi} \int d\omega \;
{\cal S}_{x_i x_j}(\omega) \,.
\end{equation}
For an initial coherent spin state ($e^{-2r}=1$), the ME
collision process does not change the collective spin variances,
but it affects the noise spectra. The $\delta$-shaped atomic spectra
of the two spins in absence of ME collisions acquire a width of
order $\gamma_{\rm exc}(N+3n)$, that is, of order $\gamma_m$. The
contribution to the total variance of the ``broad" part of the
spectrum which is not sensitive to initial squeezing in the
system, is large for the metastable state and small for the ground
state.

\section{The model for squeezing transfer}
\label{sec:model}

\begin{figure}[htb]
\centerline{\includegraphics[width=7cm,clip=]{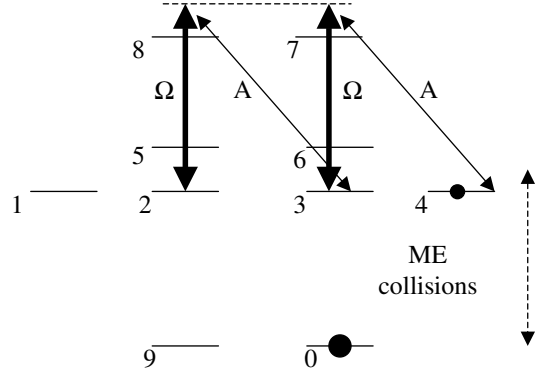}}
\caption{a) Metastable and excited sublevels of $^3$He. Three
coupling constants to the light are introduced. b) Squeezing
transfer scheme using a control field $\Omega$, a squeezed vacuum
field $A$ and metastability exchange collisions. $9$ and $0$ are
the ground state sublevels.} \label{fig:schemaHe3}
\end{figure}

In figure \ref{fig:schemaHe3} are represented the $^3$He energy
levels which are relevant for our squeezing transfer scheme. The
atoms interact with a coherent control field of Rabi frequency
$\Omega$ and frequency $\omega_1$ that we treat classically, and
a cavity field described by operators $A$ and $A^\dagger$. The
field injected into the cavity, $A_{in}$ with frequency
$\omega_2$, is in an amplitude-squeezed vacuum state: $\langle
A_{in} \rangle=0$ and $\Delta X_{in}^2=e^{-2r}$, $\Delta
Y_{in}^2=e^{2r}$, where we have introduced the field quadratures
\begin{equation}
X=A+A^\dagger \hspace{2cm} Y=i(A^\dagger - A) \,.
\end{equation}
The coherent field ($\pi$-polarized) and the squeezed vacuum
($\sigma^-$-polarized) are tuned to the blue side of the so-called
$C_9$ transition ($\lambda=1.08$ $\mu$m) from the $F=3/2$ level of
the 2$^3S$ metastable state to the 2$^3P_0$ state, the highest in
energy of the 2$^3P$ multiplicity \cite{footnote}. 
The atom-field Hamiltonian of
the system is:
\begin{equation}
H=H_0 + \hbar \left\{\Omega e^{-i\omega_1 t} (S_{73} + S_{82})
 + A( g_A S_{74} + g_B S_{83}) + h.c.\right\}
\end{equation}
where $H_0$ describes the atom-field free evolution,
$g_{A,(B)}=d_{A,(B)}\sqrt{2 \pi \omega_2/\hbar V}$ are the
coupling constants between the atoms and the cavity field, $V$
being the volume of the cavity mode and $d_{A,(B)}$ the atomic
dipoles of the transitions $7 \leftrightarrow 4$, ($8
\leftrightarrow 3$). The system is initially prepared in the fully
polarized state $\langle I_{00} \rangle=N$ and $\langle S_{44}
\rangle=n$ by preliminary optical pumping.

Non-Hamiltonian contributions to the evolution of the system
operators describe damping of the cavity mode, spontaneous
emission from the excited state and the ME collisions described in
detail in the previous section.
Linearizing the equations in the rotating frame around the fully
polarized state solution we obtain the following closed set of
equations:
\begin{eqnarray}
\dot{S}_{21}&=&-(\gamma_m- i\delta_{12}) S_{21}
    + i \Omega S_{81} + f_{21} \label{eq:uno} \\
\dot{S}_{81}&=&-(\gamma -i(\Delta_{18}-2\delta_{las})) S_{81}
    + i \Omega S_{21} + f_{81} \\
\dot{S}_{32}&=&\frac{2}{9} \gamma_m \left(-\frac{7}{2} S_{32} +
    \sqrt{3} S_{21} + S_{65} \right)+i \delta_{23} S_{32}
    -i \Omega( S_{38}-S_{72})+ f_{32} \\
\dot{S}_{72}&=&-(\gamma -i(\Delta_{27}-2\delta_{las})) S_{72}
        -i \Omega (S_{78}-S_{32})+f_{72}\\
\dot{S}_{43}&=&\gamma_m\left(-\frac{1}{3} S_{43}
    + \frac{2 \sqrt{3}}{9} (S_{32} + S_{65}) \right) +
    \frac{\sqrt{3}}{3} \gamma_f I_{09}
    + i \delta_{34} S_{43} -i \Omega S_{47} +f_{43} \\
\dot{S}_{65}&=&-\frac{7}{9}\gamma_m \left(S_{65}-\frac{2}{7}\sqrt{3} S_{21}
-\frac{2}{7} S_{32}\right) +i \delta_{56} S_{65} +f_{65} \\
\dot{S}_{47}&=&-(\gamma+i\Delta_{47}) S_{47} -i g_A n A - i \Omega
S_{43}
+f_{47} \\
\dot{S}_{38}&=&-(\gamma+i \Delta_{38}) S_{38}
    -i \Omega (S_{32}-S_{78})+f_{38} \\
\dot{S}_{78}&=&-(2\gamma- i \delta_{87}) S_{78} -i \Omega
(S_{72}-S_{38})+f_{78} \\
\dot{I}_{09}&=&\frac{1}{3}\left[-3 \gamma_f I_{09} +2\gamma_m
\left( S_{32}-\frac{1}{2} S_{65}+\frac{\sqrt{3}}{2}S_{43}+
\frac{\sqrt{3}}{2} S_{21} \right)\right] +i \delta_{90} {I}_{09} +f_{09} \\
\dot{A}&=&-(\kappa+i\Delta_C)A - i g_B S_{38} -i g_A S_{47}
    +\sqrt{2 \kappa} A_{in} \label{eq:undici}
\end{eqnarray}
where
\begin{eqnarray}
\Delta_{ij}&=&(E_{j}-E_{i})-\omega_2 \\
\delta_{ij}&=&(E_j-E_i)-\delta_{las} \\
\delta_{las}&=&\omega_1-\omega_2 \,,
\end{eqnarray}
$\gamma$ is the coherence decay rate due to spontaneous
emission from the excited state and collisions and we supposed 
$\Omega$ to be real. The non-zero
atomic diffusion coefficients are
\begin{equation}
D_{43,34}=\frac{2}{3}\gamma_m n, \hspace{0.2cm}
D_{43,90}=D_{09,34}=-\frac{2\sqrt{3}}{3}\gamma_m n, \hspace{0.2cm}
D_{47,74}= 2 \gamma n,
\hspace{0.2cm} D_{09,09}= 2 \gamma_m n \label{eq:diff}
\end{equation}
We notice that metastable variables $S_{21}$, $S_{81}$, $S_{32}$,
$S_{72}$, $S_{65}$, $S_{38}$ and $S_{78}$ form a closed subset
of equations involving Langevin forces which do not give rise to non-zero
diffusion coefficients in the fully polarized limit we consider here.
Using the same argument as in section \ref{sec:exchange},
we deduce that these variables do not
contribute to the spin noise and can be neglected. One is then
left with only four relevant equations
\begin{eqnarray}
\dot{S}_{43}&=&- \frac{\gamma_m}{3} S_{43}
    + \frac{\sqrt{3}}{3} \gamma_f I_{09}
    + i \delta_{34} S_{43} -i \Omega S_{47} +f_{43} \label{eq:S43} \\
\dot{S}_{47}&=&-(\gamma+i\Delta) S_{47} -i g_A n A
    - i \Omega S_{43} +f_{47} \label{eq:S47} \\
\dot{I}_{09}&=&\frac{1}{3}\left(-3 \gamma_f I_{09} +\gamma_m
 \sqrt{3} S_{43} \right) +i \delta_{I} {I}_{09} +f_{09}  \\
\dot{A}&=&-(\kappa+i\Delta_C)A  -i g_A S_{47} + \sqrt{2 \kappa}
A_{in} \label{eq:A}
\end{eqnarray} with $\Delta=\Delta_{47}$ and $\delta_I=\delta_{90}$.

\section{Numerical results}
\label{sec:numerics}

\begin{figure}[htb]
\centerline{\includegraphics[width=8cm,clip=]{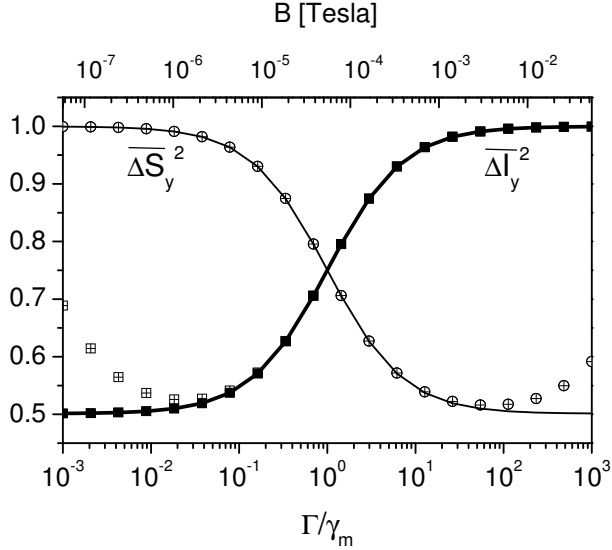}}
\caption{Symbols: numerical calculations for spin variances in
ground state (squares) and metastable state (circles), as a
function of the ratio $\Gamma/\gamma_{m}$ (lower $x$-axis). The
corresponding magnetic field needed to satisfy the resonance
conditions (\ref{eq:matchmet}) and (\ref{eq:matchfond}) is shown
in the upper $x$-axis. Numerical values of parameters are
$e^{-2r}=0.5$, $C=500$, $\kappa=100\gamma$, $\Delta=-2000 \gamma$,
$\gamma=2\times10^7$ s$^{-1}$, $\gamma_m=5\times10^6$ s$^{-1}$.
The crossed squares correspond to a calculation including an extra
relaxation rate $\gamma_0=10^{3}$ s$^{-1}$ for the metastable
variables. The lines correspond to the analytical predictions
(\ref{eq:analyI}) and (\ref{eq:analyS}). } \label{fig:num}
\end{figure}

Equations (\ref{eq:S43})-(\ref{eq:A}) can be used to find the
variances of the metastable and ground state spin numerically. A
typical result is displayed in figure \ref{fig:num}, for which we
assume that a squeezed vacuum field with $\Delta X_{in}^2=0.5$ is
injected into the cavity with the coherent control field in the
squeezing-transfer configuration.
In this figure $\overline{\Delta {S}_y^2}$ and $\overline{\Delta I_y^2}$
represent the variances of $S_y$ and $I_y$, both normalized to
their coherent spin state values. They are plotted as a function
of the ratio $\Gamma/\gamma_m$, where $\Gamma$ is the pumping
parameter
\begin{equation}
\Gamma=\gamma{3 \Omega}^2(1+C)/\Delta^2 \,, \label{eq:Gamma}
\end{equation}
and $C=g^2 n/(\kappa \gamma)$ the cooperativity. It is precisely
this ratio $\Gamma/\gamma_m$ which acts as a control parameter to
decide how the available squeezing of the field is shared
between the metastable and the ground state spin. If $\Gamma \gg
\gamma_m$, correlations are established among the metastable-state
spins, the leakage of correlation towards the ground state being
negligible. The metastable collective spin is squeezed while the
ground state spin remains unsqueezed. In the opposite limit
$\Gamma \ll \gamma_m$, spin exchange is the dominant process for
metastable atoms; they transfer their correlations to the ground
state which then becomes squeezed, while the metastable state
remains unsqueezed.

In this plot we have chosen the best conditions for squeezing
transfer:
\begin{enumerate}
\item The metastable coherence $S_{43}$ is resonantly excited
by the two fields in a Raman configuration. By introducing the
effective two-photon detuning for this coherence
\begin{equation}
\tilde{\delta}=\delta_{34}+\Omega^2/\Delta
\end{equation}
accounting for the light-shift of level 3, this condition reads 
$\tilde{\delta}=0$, or
\begin{equation}
(E_4-E_3)/\hbar+\Omega^2/\Delta=\omega_1-\omega_2 \label{eq:matchmet} \,.
\end{equation}
\item The ground state coherence $I_{09}$ should be resonantly excited
by the metastable coherence ($\delta_I=0$), i.e.
\begin{equation}
(E_0-E_9)/\hbar=\omega_1-\omega_2 \, \label{eq:matchfond} \,.
\end{equation}
\end{enumerate}

In practice a magnetic field (shown as the upper $x$-axis) can be
used to simultaneously fulfill (\ref{eq:matchmet}) and
(\ref{eq:matchfond}). When the resonance conditions are fulfilled
the difference in the Larmor frequencies in the metastable and in
the ground state is exactly compensated by the light-shift induced
by the coherent control field. Choosing $\Gamma=0.1\gamma_m$ as a
working point, the required field is about $B=57$ mG,
corresponding to $\omega_I=184$ Hz.

The vapor parameters in the figure correspond to a 1 torr sample
at 300 K, with $\gamma_m=5\times10^{6}$ s$^{-1}$ and
$\gamma=2\times10^7$ s$^{-1}$, and a metastable atom density of
$3.2 \times 10^{10}$ atoms/cm$^3$ which gives $n/N=10^{-6}$. The
symbols with a cross are a second calculation in which we added a
finite relaxation rate in the metastable state $\gamma_0$, to
account for the fact that metastable atoms are destroyed as they
reach the cell walls. We notice that only the ground state spin
squeezing in the region $\Gamma \ll \gamma_m$ is affected.

\section{Analytical results}
\label{sec:analy}

In order to have a better physical insight it is possible to find
simple analytical results within some reasonable approximation. By
adiabatic elimination of the polarization $S_{47}$ and the cavity
field assuming $\gamma, \kappa \gg \gamma_m, \gamma_f$, one
obtains
\begin{eqnarray}
\dot{{S}}_{43} + (\frac{\gamma_m}{3} +
\frac{\Gamma}{3}  -i \tilde{\delta} ) {S}_{43} & = &
\frac{\gamma_f}{\sqrt{3}} {I}_{09} + f_{43} -\frac{\Omega}{\Delta} f_{47}
     + i \frac{\Omega g n }{\Delta} \sqrt{\frac{2}{\kappa}} A_{in}
\label{eq:adiabS43} \\
\dot{{I}}_{09} + (\gamma_{f}- i \delta_{I}) {I}_{09} &=&
\frac{\gamma_m}{\sqrt{3}} {S}_{43} + f_{09}
\end{eqnarray}
In deriving (\ref{eq:adiabS43}) we assumed a Raman configuration
$\Delta \gg \gamma$, $\frac{C\gamma}{\Delta}\ll 1$ and that the
cavity detuning exactly compensates the cavity field dephasing due
to the atoms: $\Delta_C=C\kappa \gamma/\Delta$. From equation
(\ref{eq:adiabS43}) we see that $(\gamma_m+\Gamma)/3$ is the
inverse of the characteristic time constant for the metastable
coherence evolution.

\subsection{Resonant case}
If the resonance conditions (\ref{eq:matchmet}) and
(\ref{eq:matchfond}) are satisfied
($\tilde{\delta}=\delta_{I}=0$) and in the limit $\gamma_f \ll
\Gamma, \gamma_m$, we can calculate the variances of the
metastable and ground state spins
\begin{eqnarray} \Delta I_y^2 &=& \frac{N}{4} \left\{
1 - \frac{\gamma_m}{\Gamma+\gamma_m}
    \frac{C}{C+1} (1-e^{-2r}) \right\} \label{eq:analyI} \\
\Delta S_y^2 &=& \frac{n}{4} \left\{
1 - \frac{\Gamma}{\Gamma+\gamma_m}
    \frac{C}{C+1} (1-e^{-2r}) \right\}
        \label{eq:analyS}
\end{eqnarray}
which are plotted as full lines in figure \ref{fig:num}.

\subsection{Non-perfectly resonant case}
In order to test the robustness of our scheme, let us now
concentrate on what happens if the resonance conditions
(\ref{eq:matchmet}) and (\ref{eq:matchfond}) are only
approximatively satisfied. We will focus on the variance of the
ground state spin coherence $I_{09}$.

By adiabatically eliminating the metastable coherence $S_{43}$ one obtains
\begin{equation}
\dot{{I}}_{09} + \left[ \Gamma_F + i b \right] {I}_{09}
 =  f_{09} + \frac{\gamma_m\sqrt{3}}{\gamma_m+\Gamma-i 3\tilde{\delta}}
\left(f_{43} -\frac{ \Omega}{\Delta} f_{47}
  + i \frac{ \Omega g n }{\Delta} \sqrt{\frac{2}{\kappa}} A_{in} \right)
\label{eq:I09}
\end{equation}
The real part in the brackets
\begin{equation}
\Gamma_F=\gamma_f \frac{\Gamma (\gamma_m + \Gamma)+ (3 \tilde{\delta})^2}
     {(\gamma_m + \Gamma)^2+(3 \tilde{\delta})^2}
\end{equation}
is the inverse of the effective time constant for the ground state
coherence evolution which would also be the ``writing" (or
``reading") time of the quantum memory. $\Gamma_F^{-1}=2 s$ in the
example of figure \ref{fig:num} for $\Gamma=0.1\gamma_m$. It would
be proportionally shortened by increasing the metastable atoms
density although Penning collisions prevent in practice metastable
atoms densities exceeding $10^{10}$-$10^{11}$ at/cm$^2$. The
imaginary part in the brackets
\begin{equation}
b=-\left( \gamma_f \frac{3\tilde{\delta}\gamma_m}{(\gamma_m +
\Gamma)^2+(3 \tilde{\delta})^2} + \delta_{I} \right)
\end{equation}
is a light-shift ``brought back" to the ground state, which is
zero in the resonant case. Equation (\ref{eq:I09}) can be used to
calculate the best squeezing (optimized with respect to the
transverse spin quadrature) of the ground state coherence: $\Delta
I_{best}^2 = \min_{\theta} \Delta I_{\theta}^2$ with $I_{\theta}=
I_x\cos\theta+I_y\sin\theta$. We obtain
\begin{equation}
\Delta I_{best}^2 = \frac{N}{4} \left\{
1 - \frac{\gamma_m}{\Gamma+\gamma_m+(3\tilde{\delta})^2/\Gamma} \:
    \frac{C}{C+1} \, \left[ 1-(e^{-2r}+ m\sinh(2r) )\right] \right\}
\label{eq:analy_mism}
\end{equation}
where
\begin{equation}
m=1-\sqrt{\frac{1}{1+(b/\Gamma_F)^2}}
\end{equation}

\begin{figure}[htb]
\centerline{\includegraphics[width=10cm,clip=]{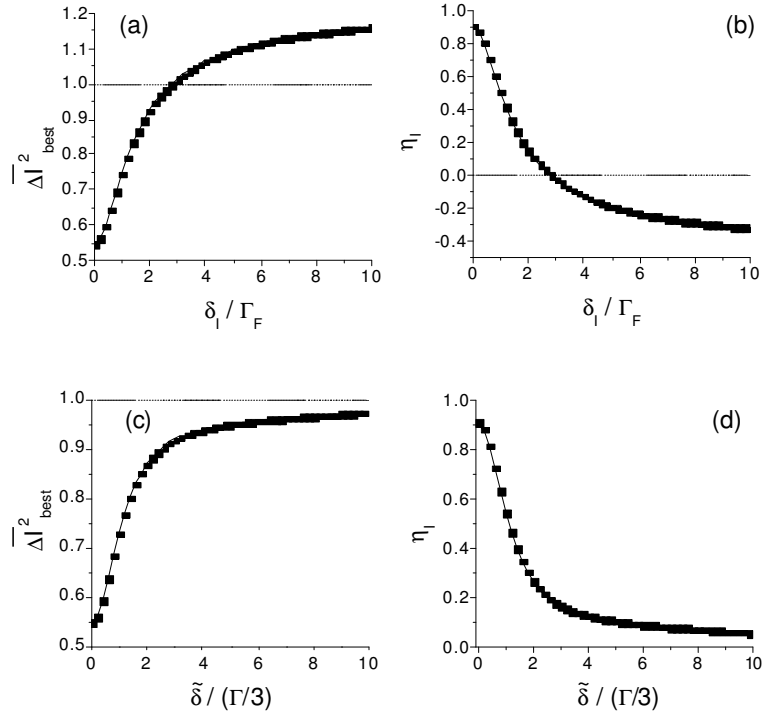}}
\caption{ Normalized ground state spin variance 
corresponding quantum transfer efficiency $\eta_I$ as a function
of $\delta_I/\Gamma_F$ (while $\tilde{\delta}=0)$ (plots (a) and (b)), or
$\tilde{\delta}/(\Gamma/3)$ (while $\delta_I=0)$ (plots (c) and (d). 
Symbols: numerical integration of equations
(\ref{eq:uno})-(\ref{eq:undici}). Lines: analytical expression
(\ref{eq:analy_mism}). Parameters are the same as in figure
\ref{fig:num} and $\Gamma=0.1 \gamma_m$.} \label{fig:mismatch2N}
\end{figure}

\begin{figure}[htb]
\centerline{\includegraphics[width=11cm,clip=]{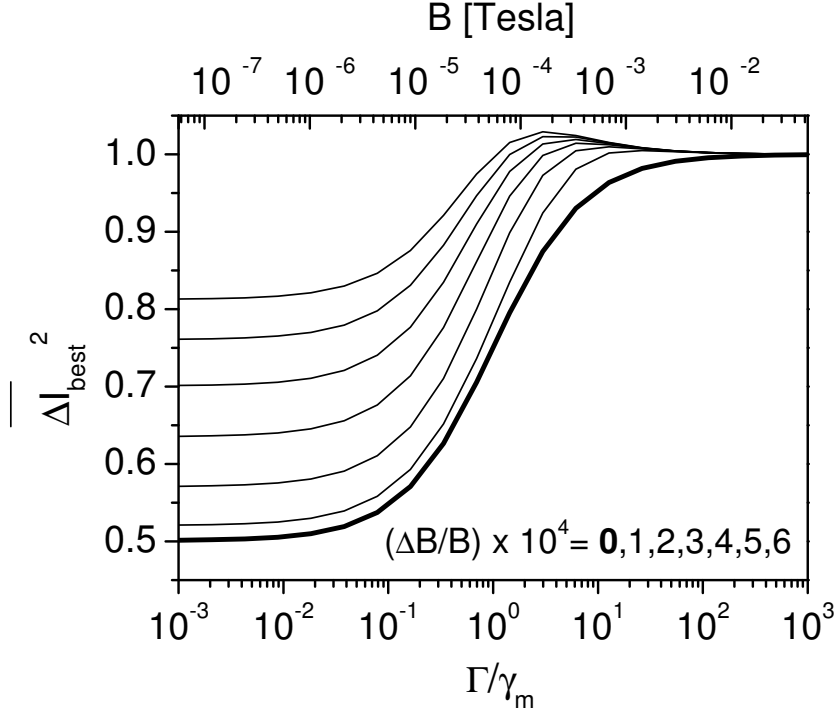}}
\caption{Normalized ``best" variance of the ground state spin as a
function of $\Gamma/\gamma_m$ (lower $x$-axis) for an increasing
inhomogeneity $\Delta B/B$ from zero to $6\times 10^{-4}$ by steps
of $1\times 10^{-4}$. On the upper $x$-axis we show the
corresponding homogeneous magnetic field needed to satisfy
resonance conditions (\ref{eq:matchmet}) and (\ref{eq:matchfond}).
Numerical values of parameters are $e^{-2r}=0.5$, $C=500$,
$\kappa=100\gamma$, $\Delta=-2000 \gamma$, $\gamma=2\times10^7$
s$^{-1}$, $\gamma_m=5\times10^6$ s$^{-1}$, $\gamma_0=0$. }
\label{fig:Bhomo}
\end{figure}

We show in figure \ref{fig:mismatch2N} the effect of a frequency
mismatch in on the normalized spin variance,
$\overline{\Delta {I}_y^2}$ and the corresponding squeezing transfer
efficiency $\eta_I$
\begin{equation}
\eta_I = \frac{1-\overline{\Delta {I}_y^2}}{1-e^{-2r}} \,.
\end{equation}
In this example, a frequency mismatch of the order of $\Gamma/3$
in the metastable state or of the order of $\Gamma_F$ in the
ground state affects the efficiency of the squeezing transfer. The
condition for the ground state frequency matching
(\ref{eq:matchfond}) imposes stringent requirements on the
homogeneity of the magnetic field. Because of the $\sinh(2r)$ in
equation (\ref{eq:analy_mism}), the larger the squeezing the worse
are the consequences of a mismatch in the condition on
$\delta_I=0$ on the ground state atoms. Physically, if a
significant dephasing between the squeezed field and the ground
state coherence builds up during the squeezing transfer time, the
atoms will see an average between the squeezed and the
anti-squeezed quadrature of the field noise. We can easily
estimate the required magnetic field homogeneity as follows. Let
us introduce the Larmor evolution frequencies in the metastable
and ground states: in low field, $\hbar \omega_\alpha=\mu_\alpha
B$ ($\alpha$=I,S) with $\mu_I/h=3.24$kHz/G and
$\mu_S/h=1.87$MHz/G, and let $\Delta B$ be the maximum field
difference with respect to the optimal value in the cell volume.
For low field, the condition on $\Delta B$ to preserve the
transfer efficiency reads $\mu_I \Delta B \ll h \Gamma_F$. Since
$\Omega^2/\Delta \simeq \Gamma \frac{\Delta}{3 \gamma C} \simeq
\frac{\mu_S}{h} B$ we get
$\frac{\Gamma}{\Gamma_F}\frac{\mu_I}{\mu_S}\frac{\Delta}{3 \gamma
C} \frac{\Delta B}{B}<1$ or, in the regime $\Gamma \ll \gamma_m$,
$600 \frac{\Delta}{\gamma C} \frac{\Delta B}{B} \ll 1$. With the
parameters of figure \ref{fig:num} this gives a condition on the
magnetic field inhomogeneity: $\Delta B/B \ll 4 \times 10^{-4}$.
In figure \ref{fig:Bhomo} we calculated the variance of the ground
state spin as a function of $\Gamma/\gamma_m$ for an increasing
inhomogeneity $\Delta B/B$ from zero (thick line) to $6\times
10^{-4}$. In practice a homogeneity of 100 ppm should be
sufficient for the chosen parameters to guarantee that all atoms
will be squeezed.

\section{Optical readout}
\label{sec:lecture}

\subsection{Outgoing field squeezing}

As briefly stressed in \cite{prl} the squeezed fluctuations which
are stored into the nuclear spins can be retrieved optically in
the field exiting the cavity by using the reverse transfer
process. Indeed, once the write sequence of the quantum memory has
been completed, both the fields and the discharge can be switched
off, leaving the atoms in the fundamental state in a spin-squeezed
state. After a variable storage time, switching back on 
the discharge and only the control field in the same configuration 
as for the writing phase ($\Gamma\ll\gamma_m$),
will rapidly put a small
fraction of atoms in the metastable state and start the reverse
transfer process from the fundamental atoms to the field. The
correlations in the ground state will slowly transfer via the
metastable state to the intracavity field. This will then result
in squeezed fluctuations for the field exiting the cavity, which
can be measured by homodyne detection.

More quantitatively, if we still assume that the metastable spin
observables and the intra-cavity field adiabatically follow the
ground state spin observables and the evolution equations for the
fluctuations of the squeezed component are in the resonant
situation, we have 
\begin{eqnarray}\label{eq:Iyt} \dot{I}_y(t)&=&-\Gamma_F I_y(t)+\beta
X_{in}(t)+\tilde{f}_y(t)\\
X_{out}(t)&=&\sqrt{2\kappa}X(t)-X_{in}(t)\label{eq:Xoutt}\end{eqnarray}
with
\begin{eqnarray}
\beta &=&\frac{\gamma_m}{\gamma_m+\Gamma}
\frac{g_A n\Omega\sqrt{3}}{2\Delta}\sqrt{\frac{2}{\kappa}}\\
\tilde{f}_y&=&\frac{\gamma_m\sqrt{3}}{\gamma_m+\Gamma}
\left[\frac{f_{43}-f_{34}}{2i}-\frac{\Omega}{\Delta}
\frac{f_{74}-f_{47}}{2i}\right] + \frac{f_{09}-f_{90}}{2i}
\end{eqnarray}
Denoting by $t=0$ the start of the readout sequence and by
$e^{-2r}=\Delta I_y^2(0)/(N/4)$ the initial squeezing of the
ground state nuclear spin, the two-time correlation function of
the outgoing field amplitude quadrature can be obtained via
(\ref{eq:Xoutt}) after integration of (\ref{eq:Iyt})
\begin{eqnarray} \mathcal{C}(t,t')\equiv
\langle X^{out}(t)
X^{out}(t')\rangle=\delta(t-t')-2\Gamma_F
\eta_I[1-e^{-2r}]e^{-\Gamma_F(t+t')}\label{eq:correl}\end{eqnarray}
The $\delta$-correlated term corresponds to the vacuum
fluctuations contribution, whereas the second term corresponds to
a transient squeezing for the outgoing field which is proportional
to the initial atomic squeezing. In (\ref{eq:correl}), $\eta_I$
designates the optimal quantum transfer efficiency in the ground
state
\begin{equation}
\label{eq:etaI}\eta_I=
\frac{\gamma_m}{\gamma_m+\Gamma}\frac{C}{C+1}
\end{equation} 
The ground state squeezing can be adequately measured by homodyne
detection using a temporally matched local oscillator as shown in
Refs.~\cite{dantanpra04,dantanpulse}. Using a local oscillator
with envelope $\mathcal{E}(t)$ the normalized power measured by a
Fourier-limited spectrum analyzer integrating over a time $T$ is
given by
\begin{eqnarray}P(t)&=&\frac{1}{\mathcal{E}(t)^2}\int_{-\pi/T}^{\pi/T}\frac{d\omega}{2\pi}
\int_{t}^{t+T}d\tau\int_t^{t+T}d\tau'e^{-i\omega(\tau-\tau')}
\mathcal{E}(\tau)\mathcal{E}(\tau')\mathcal{C}(\tau,\tau')\end{eqnarray}
In order to measure the atomic squeezing one has to maximize the
temporal overlap between the local oscillator and the field
radiated by the atoms: $\mathcal{E}(t)\propto e^{-\Gamma_Ft}$. 
For such a local oscillator and for an integration time longer than the
readout time $\Gamma_F^{-1}$ the measured power can be written as
the sum of a shot-noise term $\mathcal{N}$ and a time-dependent
signal term $\mathcal{S}$ proportional to the initial squeezing:
\begin{equation}
P(t)=\mathcal{N}-\mathcal{S}[1-e^{-2r}]e^{-2\Gamma_Ft}\end{equation}
with $\mathcal{S}\simeq \eta_I \mathcal{N}$. The ground state
nuclear spin fluctuations can therefore be measured optically with
the same efficiency $\eta_I$ as in the write sequence. However,
because of the slow character of the correlation transfer process
in the ground state the readout time is as long as the write time.
As expected it is not possible to access the quantum memory faster
during the readout than during the write phase. One could think of
a faster readout method by transferring the fundamental atoms
fluctuations to the metastable atoms and perform the optical
readout in the regime $\Gamma\gg\gamma_m$. However, as we showed
in section \ref{sec:exchange}, starting with a squeezed
fundamental spin and first switching on the discharge (without the
fields) will transfer very few correlations from the fundamental
to the metastable atoms and almost no squeezing
will be retrieved in the field.

\section{Entangling two separate samples}
\label{sec:2cells}

A direct and important extension of the previous results is that it
is possible to transfer quantum correlations between different light
beams to two spatially separated nuclear spins. If one disposes of
EPR fields this allows to entangle two separate ensembles
\cite{dantanEPL04}. Such EPR atomic states are very useful for
quantum information protocols involving the manipulation of
continuous variable entanglement, such as atomic teleportation for
instance \cite{dantanprl05}.

Let us consider two identical ensembles 1 and 2 illuminated by
EPR-correlated vacuum fields $A^{(i)}_{in}$ and coherent control
fields $\Omega_i$ ($i=1,2$). Without loss of generality we assume
symmetrical field correlations of the form
\begin{eqnarray}\Delta^2 X^{(i)}_{in}&=&\Delta^2 Y^{(i)}_{in}=\cosh(2r)\hspace{0.3cm}(i=1,2)\\
\langle X^{(i)}_{in} X^{(j)}_{in}\rangle &=&-\langle Y^{(i)}_{in}
Y^{(j)}_{in}\rangle=\sinh(2r)\hspace{0.3cm}(i\neq
j),\end{eqnarray} i.e. that the amplitude quadratures are
correlated and the phase quadratures anti-correlated:
$\Delta^2(X^{(1)}_{in}-X^{(2)}_{in})=
\Delta^2(Y^{(1)}_{in}+Y^{(2)}_{in})=2e^{-2r}$.
For perfect entanglement ($r=\infty$) these EPR variances vanish.
Both spins are initially prepared in a coherent spin state and we
assume an equal incident power on both samples
($\Omega_1=\Omega_2$). Under the same adiabatic approximations as
before, the fluctuations of the transverse spin components satisfy
equation of the form (\ref{eq:Iyt})
\begin{eqnarray} \dot{I}_{xi}&=&-\Gamma_F
I_{xi}-\beta
Y^{(i)}_{in}+\tilde{f}_{xi},\\
\dot{I}_{yi}&=&-\Gamma_F I_{yi}+\beta
X^{(i)}_{in}+\tilde{f}_{yi},\end{eqnarray} ($i=1,2$). Because of
the linearity of the coupling in this regime, the EPR atomic
nuclear spin operators, $I_{x1}+I_{x2}$ and $I_{y1}-I_{y2}$, are
clearly coupled to the EPR field operators \begin{eqnarray}
\frac{d}{dt}(I_{x1}+I_{x2})&=&-\Gamma_F(I_{x1}+I_{x2})-\beta(Y^{(1)}_{in}+Y^{(2)}_{in})
+\tilde{f}_{x1}+\tilde{f}_{x2}\label{eq:sommeIx}\\
\frac{d}{dt}(I_{y1}-I_{y2})&=&-\Gamma_F(I_{y1}-I_{y2})+\beta(X^{(1)}_{in}-X^{(2)}_{in})
+\tilde{f}_{y1}-\tilde{f}_{y2}\label{eq:diffIy}\end{eqnarray} The
amount of EPR-type correlations between the incident fields is
given by the half-sum of the EPR variances
\begin{equation}\mathcal{E}_f=\frac{1}{2}\left[\Delta^2(X^{(1)}_{in}-X^{(2)}_{in})+\Delta^2(Y^{(1)}_{in}+Y^{(2)}_{in})\right]=2e^{-2r}\end{equation}
In the Gaussian approximation the entanglement between the nuclear
spins can be evaluated using the same quantity (also normalized to
2)
\begin{equation}
\mathcal{E}_I=\frac{2}{N}\left[\Delta^2(I_{x1}+I_{x2})+\Delta^2(I_{y1}-I_{y2})\right]\end{equation}
It follows from (\ref{eq:sommeIx}-\ref{eq:diffIy}) that the last
two quantities are simply related by
\begin{equation}\mathcal{E}_I=\eta_I\mathcal{E}_f+2(1-\eta_I).\end{equation}
Like squeezing entanglement can also be in principle perfectly
mapped onto the nuclear spins with an efficiency $\eta_I$
(\ref{eq:etaI}), close to unity in the regime $\Gamma\gg\gamma_m$
and $C\gg 1$. 
Let us introduce the correlation functions $\mathcal{C}_I^{(i,i)}$
of individual spins \textit{inside} the ensemble $i$ (i=1,2):
\begin{equation}
\mathcal{C}_I^{(i,i)}=\frac{ \overline{\Delta I_{xi}^2}  -1}{4N}
\hspace{1cm} (i=1,2)
\end{equation}
and the correlation function $\mathcal{C}_I^{(i,j)}$ 
of two individual spins belonging to the \textit{different} ensembles
$i$ and $j$:
\begin{equation}
\mathcal{C}_I^{(i,j)}=\frac{\overline{\langle I_{xi} I_{xj} \rangle} }{4N}
\hspace{1cm} (i \neq j = 1,2)
\end{equation}
where the overline indicates the normalization of the correlation functions
to $N/4$. In our case for $\eta_I \simeq 1$ we get:
\begin{eqnarray}
\mathcal{C}_I^{(1,1)}&=&\mathcal{C}_I^{(2,2)} \simeq \frac{\cosh(2r)-1}{4N} \\
\mathcal{C}_I^{(1,2)}&=&\mathcal{C}_I^{(2,1)} \simeq \frac{\sinh(2r)}{4N} \,.
\end{eqnarray}
It is interesting to note that the two correlation functions 
$\mathcal{C}_I^{(1,1)}$ and $\mathcal{C}_I^{(1,2)}$
become approximately equal for a large entanglement $e^{2r} \gg 1$
so that an individual spin is about as much correlated with 
the other spins in its own ensemble as with the spins of the other ensemble.

\section{The imperfect polarization case}
\label{sec:Pdiff1}
\begin{figure}[htb]
\centerline{\includegraphics[width=5cm,clip=]{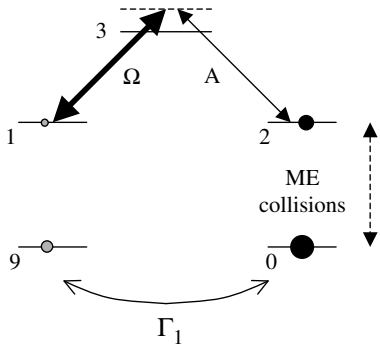}}
\caption{Sublevels $1$ and $2$ are metastable, level $3$ is the
excited state, $9$ and $0$ are the ground state sublevels. We include
a relaxation process in the ground state so that $P \neq 1$ in steady state.}
\label{fig:schema2N}
\end{figure}

The nuclear polarization of the sample is defined as
\begin{equation}
P=\frac{\langle I_{00}\rangle - \langle I_{99}\rangle}
       {\langle I_{00}\rangle + \langle I_{99}\rangle}
\end{equation}
In practice polarization between 80\% and 85\% are currently
achieved by optical pumping in dilute $^3$He samples
\cite{nacher85}. If the atoms are prepared in a state which is not
fully polarized - $P \neq 1$ - the situation is clearly more
complicated than we described in \cite{prl} and in the present
paper. In particular, equations (\ref{eq:uno})-(\ref{eq:undici})
and (\ref{eq:diff}) obtained by linearization around the fully
polarized state are no longer valid. We did not perform a complete
analysis in the $P \neq 1$ case. However, one can have a good idea
of the result by using the simplified model of \cite{prl} which
involves only two metastable sublevels (see figure
\ref{fig:schema2N}). As in section \ref{sec:model}, a Raman
transition is driven by a coherent control field of Rabi frequency
$\Omega$ and a squeezed vacuum cavity field:
\begin{equation}
H=H_0 + \hbar \left\{ \Omega \, \tilde{S}_{31}e^{-i \omega_1 t} +
 g \, A  \, \tilde{S}_{32} + \rm{h.c.} \right\} \,.
\end{equation}
In this toy-model the control field $\Omega$ also acts as an
optical pumping beam (able to transfer the atoms from sublevel $9$
to sublevel $0$) and we introduce explicitly a relaxation in the
ground state, so that $P \neq 1$ in steady state.
Let us introduce for this model the rescaled coupling constant $\tilde{g}$,
the atomic one-photon detunings $\Delta_1$ and $\Delta$, the
two-photon detunings $\tilde \delta$ and $\delta_I$,
and two pumping parameters $\Gamma_p$ and $\Gamma$:
\begin{eqnarray}
\Delta_1&=&(E_3-E_1)/\hbar-\omega_1, \hspace{1cm}
\Delta=(E_3-E_2)/\hbar-\omega_2, \\
\tilde{\delta}&=&\Delta_1-\Delta + \frac{\Omega^2}{\Delta},
\hspace{2cm}
\delta_I=(E_0-E_9)/\hbar-(\omega_1-\omega_2),\\
\Gamma_P&=&\gamma \frac{\Omega^2}{\Delta^2}, \hspace{3.5cm}
\Gamma=\Gamma_p(1+C),
\end{eqnarray}
where $\gamma$ is the optical coherence decay rate and $C$ is the
cooperativity parameter defined by equation (\ref{eq:Gamma}). For
the atomic operators we introduce $\tilde{S}_+=\tilde{S}_{21}$, $\tilde{S}_-=\tilde{S}_{12}$
\begin{equation}
\tilde{S}_x=(\tilde{S}_- + \tilde{S}_+)/2,\hspace{1cm} 
\tilde{S}_y=i(\tilde{S}_- - \tilde{S}_+)/2, \hspace{1cm}
\tilde{S}_z=(\tilde{S}_{22} - \tilde{S}_{11})/2,\hspace{1cm}
\end{equation}
and similarly for the ground state operators. In the limit of
large one photon detunings the excited state and the optical
coherences can be adiabatically eliminated, yielding a set of
equations similar to those of Ref.~\cite{dantanPRA03} with the
addition of metastability exchange. By adiabatically eliminating
the field (assumed to be resonant in the cavity) and for
$\tilde{\delta}=0$, $\delta_I=0$, we obtain:
\begin{eqnarray}
\dot{\tilde{S}}_{+}&=&-(\Gamma_p+\gamma_m) \tilde{S}_{+} + \gamma_f I_+
     + 2i \tilde{g} A \tilde{S}_z +    + f_{\tilde{S}_+} \label{eq:2NMnot1_1} \\
\dot{\tilde{S}}_{z}&=&-(\Gamma_p+\gamma_m) \tilde{S}_{z} + \gamma_f I_z
        + \frac{n \Gamma_p}{2}
        + i\tilde{g}[A^{\dagger} \tilde{S}_+ -A \tilde{S}_-]+ f_{\tilde{S}_z} \\
\dot{A}&=&-(\kappa+i\Delta_C)A+i\tilde{g}\tilde{S}_++\sqrt{2\kappa}A_{in}\\
\dot{I}_+&=&-(\gamma_f + \Gamma_1) I_+ + \gamma_m \tilde{S}_+ + f_{I_+} \\
\dot{I}_z&=&-(\gamma_f + \Gamma_1) I_z + \gamma_m \tilde{S}_z + f_{I_z}
\label{eq:2NMnot1_4}
\end{eqnarray}
The semiclassical version of equations
(\ref{eq:2NMnot1_1})-(\ref{eq:2NMnot1_4}) has a stationary
solution $\langle \tilde{S}_+ \rangle=\langle I_+ \rangle=\langle A_{in}
\rangle=0$ and with
\begin{eqnarray}
\langle \tilde{S}_z \rangle&=&
\frac{\gamma_f+\Gamma_1}{\gamma_m}\langle I_z \rangle
\hspace{2cm}
P=\frac{1}{1+\Gamma_1(\Gamma_p+\gamma_m)/(\Gamma_p \gamma_f)}
\end{eqnarray}
We will have in practice $\Gamma_1 \ll \gamma_f$, meaning that the
nuclear polarization in the metastable state $P^\ast=\langle \tilde{S}_z
\rangle/(n/2)$ and the nuclear polarization in the ground state
$P=\langle I_z \rangle/(N/2)$ are almost equal. In this toy-model
the stationary $P$ is determined by the balance between the decay
$\Gamma_1$ and the pumping $\Gamma_p$. In reality, the atoms will
be previously pumped more efficiently with resonant light. When we
linearize the equations around the steady state we obtain
\begin{eqnarray}
\dot{\tilde{S}}_+&=&-(\tilde{\Gamma}+\gamma_m)\tilde{S}_+ + \frac{i g \Omega
n}{\kappa \Delta} P^\ast \sqrt{2\kappa} A_{in} + \gamma_f I_+ +
f_{\tilde{S}_+}
\label{eq:S-}\\
\dot{I}_+ &=& -(\gamma_f+\Gamma_1) I_+ + \gamma_m \tilde{S}_+ + f_{I_+}
\label{eq:I-}
\end{eqnarray}
with
\begin{equation}
\tilde{\Gamma}=\Gamma_p(1+C P^\ast)
\end{equation}
Starting from equations (\ref{eq:S-})-(\ref{eq:I-}) one can proceed as in
section \ref{sec:analy} to obtain
\begin{equation}
\overline{\Delta {I}_y^2 }= \frac{\Delta
I_y^2}{NP/4}=\frac{1}{P}+
(P^* e^{-2r}-1)\eta_I'+\frac{\eta_I'}{2\tilde{C}}\frac{P^*-1}{P}
\end{equation}
where
\begin{eqnarray}
\tilde{C}&=&C P^\ast \hspace{1cm}\textrm{and}\hspace{1cm}
\tilde{\Gamma}_f=\frac{\gamma_f\tilde{\Gamma}}{\tilde{\Gamma}+\gamma_m}\\
\eta_I'&=&\frac{\tilde{C}}{\tilde{C}+1} \;
\frac{\gamma_m}{\tilde{\Gamma}+\gamma_m} \;
\frac{\tilde{\Gamma}_f}{\tilde{\Gamma}_f+\Gamma_1}
\end{eqnarray}
For $P^\ast \simeq P$ and $\tilde{C}\gg 1$, we have finally
\begin{equation}
\overline{\Delta {I}_y^2} = \eta_I'e^{-2r}+(1-\eta_I')/P
\label{eq:analyIPnot1}
\end{equation}
Equation (\ref{eq:analyIPnot1}) shows that the main consequence of
having $P \neq 1$ is a rescaling of the cooperativity and the
pumping parameter $\tilde{\Gamma}$ and the quantum transfer
efficiency $\eta_I'$, which are reduced by a factor $P$. Let us
note that, for $P\neq 1$, when no squeezing enters the cavity, the
atoms are no longer in a coherent spin state. This shows, however,
that strong squeezing transfer is still possible with a non-ideal
polarization.

\begin{acknowledgments} Laboratoire Kastler Brossel is UMR 8552 du
CNRS, de l'ENS et de l'UPMC. This work was supported by the
COVAQIAL European project No. FP6-511004.
\end{acknowledgments}

\section{Appendix}
Evolution equations of the density matrix elements under ME
collisions are:

\begin{eqnarray}
{\frac {d}{dt}}\rho_{{11}}= & \gamma_{{{\rm exc}}} &  \left(
-N\rho_{{11}}+\frac{ 1 }{ 3 } \rho_{{99}}  \left( \rho_ {{22}}+3
\rho_{{11}}+2 \rho_{{55}} \right) \right) \nonumber \
\label{eq:nonlinME_1}
\\
{\frac{d}{dt}}\rho_{{12}}= & \gamma_{{{\rm exc}}} &  \left(
-N\rho_{{12} }+\frac{ 2 }{ 9 } \rho_{{99}} \left(  \left(
\rho_{{23}}+\rho_{{56}} \right) \sqrt {3}+3 \rho_{{12}} \right)
+\frac{ \sqrt {3} }{ 9 } \rho_{{90}} \left( \rho_{{22}}+3 \rho_{
{11}}+2 \rho_{{55}} \right) \right) \nonumber \
\\
{\frac{d}{dt}}\rho_{{13}}= & \gamma_{{{\rm exc}}} &  \left(
-N\rho_{{13}}+\frac{ 1 }{ 3 } \rho_{{99}} \left(
\rho_{{13}}+\rho_{{24}} \right) +\frac{ 2 }{ 9 } \rho_{{90 }} \left(
\left( \rho_{{23}}+\rho_{{56}} \right) \sqrt {3}+3 \rho_ {{12}}
\right) \right) \nonumber \
\\
{\frac {d}{dt}}\rho_{{1 4}}= & \gamma_{{{\rm exc}}} &  \left(
-N\rho_{{14}}+\frac{ \sqrt {3} }{ 3 } \rho_{{90}}  \left(
\rho_{{13}}+\rho_{{24}} \right) \right) \nonumber \
\\
{\frac {d}{dt}}\rho_{{22}}= & \gamma_{{{\rm exc}}} &  \left(
-N\rho_{{22}}+\frac{ 2 }{ 9 } \rho_{{99}} \left( 2 \rho_{{22}
}+\rho_{{55}}+\rho_{{66}}+2 \rho_{{33}} \right) +\frac{ 2 }{ 9 }
\rho_{ {90}} \left( \sqrt {3}\rho_{{21}}+\rho_{{65}}+\rho_{{32}}
\right) \right. \nonumber \
\\  & & \,\,\,\,  \left.    +\frac{ 2 }{ 9 }  \rho_{{09}} \left( \sqrt
{3}\rho_{{12}}+\rho_{{23}}+\rho_{{56}} \right)+\frac{ 1 }{ 9 }
\rho_{{00}} \left( \rho_{{22}}+3 \rho_{{11}}+2 \rho_{{55}} \right)
\right) \nonumber \
\\
{\frac {d}{dt}}\rho_{{23}}= & \gamma_{{{\rm exc}}} &  \left(
-N\rho_{{23}}+\frac{ 2 }{ 9 }  \rho_{{99}} \left( \sqrt {3}\rho
_{{34}}+\rho_{{56}}+\rho_{{23}} \right)+\frac{ 2 }{ 9 } \rho_ {{90}}
\left( 2 \rho_{{22}}+\rho_{{55}}+\rho_{{66}}+2 \rho_{{33}} \right)
\right. \nonumber \
\\  & & \,\,\,\,  \left.    +\frac{ \sqrt {3} }{ 9 }  \rho_{ {09}} \left(
\rho_{{13}}+\rho_{{24}} \right)  +\frac{ 2 }{ 9 } \rho_{{00}} \left(
\sqrt {3}\rho_{{12}}+\rho_{{23}}+\rho_{{56}} \right) \right)
\nonumber \
\\
{\frac {d}{dt}}\rho_{{24}}= & \gamma_{{{\rm exc}}} &  \left(
-N\rho_{{24}}+\frac{ 2 }{ 9 } \rho_{{90}} \left( \sqrt {3} \left(
\rho_ {{23}}+\rho_{{56}} \right)+3 \rho_{{34}} \right) +\frac{ 1 }{
3 } \rho_{{00}}  \left( \rho_{{13}}+\rho_{{24}} \right) \right)
\nonumber \
\\
{\frac {d}{dt}}\rho_{{33}}= & \gamma_{{{\rm exc}}} &  \left(
-N\rho_{{33}}+\frac{ 1 }{ 9 }  \rho_{{99}} \left( 2
\rho_{{66}}+\rho_{{33}}+3 \rho_{{44}} \right)+\frac{ 2 }{ 9 }
\rho_{{90}}\left( \sqrt {3}\rho_{{43}}+\rho_{{65}}+ \rho_{{32}}
\right)  \right. \nonumber \
\\  & & \,\,\,\,  \left.    +\frac{ 2 }{ 9 } \rho_{{09}} \left( \sqrt {3}\rho_{{34}}+
\rho_{{56}}+\rho_{{23}} \right) +\frac{ 2 }{ 9 } \rho_{{00}} \left(
2 \rho_{{ 22}}+\rho_{{55}}+\rho_{{66}}+2 \rho_{{33}} \right) \right)
\nonumber \
\\
{\frac {d}{dt}}\rho_{{34}}= & \gamma_{{{\rm exc}}} &  \left(
-N\rho_{{34}}+\frac{ \sqrt {3} }{ 9 } \rho_{{90}}  \left( 2
\rho_{{66}}+\rho_{{33}}+3 \rho_{{44}} \right)+\frac{ 2 }{ 9 }
\rho_{{00}} \left( \sqrt {3} \left( \rho_{ {23}}+\rho_{{56}}
\right)+3 \rho_{{34}} \right) \right) \nonumber \
\\
{\frac {d}{dt}}\rho_{{44} }= & \gamma_{{{\rm exc}}} &  \left(
-N\rho_{{44}}+\frac{ 1 }{ 3 } \rho_{{00}}  \left( 2
\rho_{{66}}+\rho_{{33}}+3 \rho_{{44}} \right) \right) \nonumber \
\\
{\frac {d}{dt}}\rho_{{55}}= & \gamma_{{{\rm exc}}} &  \left(
-N\rho_{{55}}+\frac{ 1 }{ 9 } \rho_{{ 99}}  \left( 2
\rho_{{22}}+\rho_{{55}}+\rho_{{66}}+2 \rho_{{33}} \right)-\frac{ 2
}{ 9 } \rho_{{90}} \left( \sqrt
{3}\rho_{{21}}+\rho_{{65}}+\rho_{{32}} \right) \right. \nonumber \
\\  & & \,\,\,\,  \left.   -\frac{ 2 }{ 9 }
\rho_{{09}} \left( \sqrt {3}\rho_{{12}}+\rho_{{23}}+ \rho_{{56}}
\right) +\frac{ 2 }{ 9 } \rho_{{00}}  \left( \rho_{{22}}+3 \rho_{{
11}}+2 \rho_{{55}} \right) \right) \nonumber \
\\
{\frac{d}{dt}}\rho_{{56}}= & \gamma_{{{\rm exc}}} &  \left(
-N\rho_{{56}}+\frac{ 2 }{ 9 } \rho_{{99}}  \left( \sqrt
{3}\rho_{{34}}+\rho_{{56}}+\rho_{{23}} \right)-\frac{ 1 }{ 9 }
\rho_{{90}} \left(2 \rho_{{22}}+\rho_{{55}}+\rho_{{66}}+2
\rho_{{33}} \right) \right. \nonumber \
\\  & & \,\,\,\,  \left.   -\frac{ 2 \sqrt {3} }{ 9 }  \rho_{{09}} \left(
\rho_{{13}}+\rho_{{24}} \right)+\frac{ 2 }{ 9 } \rho_{{00}} \left(
\sqrt {3}\rho_{{12}}+\rho_{{23}}+ \rho_{{56}} \right) \right)
\nonumber \
\\
{\frac {d}{dt}}\rho_{{66}}= & \gamma_{{{\rm exc}}} &  \left(
-N\rho_{{66}}+\frac{ 2 }{ 9 } \rho_{{99}}  \left( 2 \rho_{{66
}}+\rho_{{33}}+3 \rho_{{44}} \right)-\frac{ 2 }{ 9 } \rho_{{90}}
\left( \sqrt {3}\rho_{{43}}+\rho_{{65}}+\rho_{{32}} \right)\right.
\nonumber \
\\  & & \,\,\,\,  \left.   -\frac{ 2
}{ 9 } \rho_{{09}} \left( \sqrt
{3}\rho_{{34}}+\rho_{{56}}+\rho_{{23}} \right) +\frac{ 1 }{ 9 }
\rho_{{00}} \left( 2 \rho_{{22}}+\rho_{{55}}+\rho_{{66}}+2
\rho_{{33}} \right) \right) \nonumber \
\\
{ \frac{d}{dt}}\rho_{{00}}= & \gamma_{{{\rm exc}}} &  \left(
-n\rho_{{00}}+\frac{ 1 }{ 3 }  \left( 3 \rho
_{{44}}+\rho_{{66}}+\rho_{{22}}+2 \rho_{{55}}+2 \rho_{{33}} \right)
N \right) \nonumber \
\\
{\frac {d}{dt}}\rho_{{09}}= & \gamma_{{{\rm exc}}} &  \left(
-n\rho_{{09}}+\frac{ 1 }{ 3 } N \left(  \left(
\rho_{{43}}+\rho_{{21}} \right) \sqrt {3}+2 \rho_{{32}}-\rho_{{65}}
\right)  \right) \nonumber \
\\
{\frac {d}{dt}}\rho_{{99}}= & \gamma_{{{\rm exc}}} &  \left(
-n\rho_{{99}}+\frac{ 1 }{ 3 }  \left( \rho_{{33}}+2 \rho_{{22}}+3
\rho_{{11}}+\rho_{{55}}+2 \rho_{{66}} \right) N \right) \nonumber \
\label{eq:nonlinME_last}
\end{eqnarray}


\begin{thebibliography}{99}
\bibitem{footnote}
In practice one wants the squeezed field and the coherent field
to be copropagating. The squeezed field should thus be
$\sigma$-polarized, with a $\sigma^+$ component; we
checked that this component plays no role here.
\end{thebibliography}
\end{document}